\documentclass[epj]{svjour}
\usepackage{latexsym}
\usepackage{amsmath}
\usepackage{amssymb}
\usepackage{epsfig}
\usepackage{graphicx}
\usepackage{dcolumn}
\usepackage{bm}
\usepackage{color}
\usepackage{empheq}

\newcommand{\bsigma}{\mbox{\boldmath$\sigma$}}

\begin{document}
\title{Symmetry energy in nuclear density functional theory}
\author{
W. Nazarewicz\inst{1-3} 
\and 
P.-G. Reinhard\inst{4}
\and
W. Satu{\l}a\inst{3}
\and
D. Vretenar\inst{5}
}                     
\institute{
Department of Physics and Astronomy, University of Tennessee Knoxville, Tennessee 37996, USA
\and
Oak Ridge National Laboratory, P.O. Box 2008, Oak Ridge, Tennessee 37831, USA
\and
Faculty of Physics, University of Warsaw, ul. Ho\.za 69, 00-681 Warsaw, Poland
\and
Institut f\"ur Theoretische Physik, Staudtstr. 7
D-90158 Universit\"at Erlangen/N\"urnberg, Erlangen, Germany
\and
Physics Department, Faculty of Science, University of Zagreb, Zagreb, Croatia
}
\date{Received: date / Revised version: date}
%
\abstract{The nuclear symmetry energy represents a response 
to the neutron-proton asymmetry. In this survey we discuss various aspects of
symmetry energy in the framework of nuclear density functional theory, considering
both non-relativistic and relativistic self-consistent mean-field realizations side-by-side.
Key observables pertaining to bulk nucleonic matter  and finite nuclei are reviewed. Constraints on the symmetry energy and correlations between observables and symmetry-energy parameters, using statistical covariance analysis, are investigated. Perspectives for future work are outlined in the context of ongoing experimental efforts. 
} 
\maketitle
\section{Introduction}
\label{sec:intro}

Density Functional Theory (DFT) is a universal approach
used to describe properties of complex, strongly correlated many body systems.
Originally developed in the context of many-electron systems in condensed matter physics and quantum chemistry \cite{(Hoh64),(Koh65)} (also known under the name of Kohn-Sham DFT), it is also a tool of choice in microscopic studies of complex 
heavy nuclei. The basic implementation of this framework is in terms of
self-consistent mean-field (SCMF) models \cite{(Vau72),(Neg72),(Ben03)}. 

Extending the DFT to atomic nuclei, the nuclear DFT,  is not straightforward as nuclei are self-bound, small,
superfluid aggregations of two kinds of fermions, governed by strong surface effects. Their smallness
leads to appreciable  quantal fluctuations (finite-size effects)
which are  difficult to incorporate into the energy density
functional (EDF). The lack of external binding
potential implies that the nuclear DFT must be necessarily
formulated in terms of intrinsic normal and anomalous (pairing)
densities \cite{(Mes09)}.  
A density matrix expansion of the effective interaction
suggests that, in addition to the standard
local nucleon density, superior EDFs should also include more involved 
nucleon aggregates such as the kinetic-energy density and
spin-orbit density \cite{(Vau72),(Neg72),(Ben03)}

The commonly-used single-reference SCMF  methods include the local (Skyrme),
non-local (Gogny) and covariant (relativistic) approaches \cite{(Ben03),(Lal04),(Vre05)}. 
All these approaches
are thought to be different realizations of an underlying effective field theory \cite{(Wei99)} with the ultraviolet
physics hidden in free parameters adjusted to observations.
For that reason, predictions for low-energy (infrared) physics should be fairly independent of the 
particular variant used in calculations \cite{(Pug03),(Car08),(Dru10),(Dob12)}.
The underlying EDFs
are constructed in phenomenological way, with coupling constants  optimized
to selected nuclear data and expected properties of homogeneous nuclear matter.

In practice, nuclear EDFs differ in their functional form
and are subject to different optimization strategies   causing that their predictions vary even within
a single family of EDFs. In particular, large uncertainties remain in the 
isovector channel, which is poorly constrained by experiment.
A key quantity characterizing the
  interaction in the isovector channel is the nuclear symmetry energy (NSE)
describing the static response of the nucleus to  the neutron-proton asymmetry. 

As discussed in this Topical Issue, the NSE influences a broad spectrum
of phenomena, ranging from subtle isospin mixing effects in $N\sim Z$
nuclei to particle stability of neutron-rich nuclei, to nuclear
collective modes, and to radii and masses of neutron stars.  Various
nuclear observables are sensitive probes of NSE, and numerous
phenomenological indicators can be constructed to probe its various
aspects.

It is the aim of this contribution to analyze the relations between  
NSE and measurable observables in finite nuclei. The most promising
observables for isovector properties that have stimulated
vigorous experimental and theoretical activity include neutron radii,
neutron skins, dipole polarizability, and neutron star radii.  The
ongoing efforts are focused on better constraining the uncertainties
concerning the equation of state (EOS) of the symmetric and asymmetric
nucleonic matter (NM) and, in particular, the symmetry energy and its density
dependence. Parameters that characterize the NSE
are not entirely independent. They are affected by key nuclear
observables in different ways. Thus it is {\it the sine qua non}
of a further progress in this area to understand the correlation
pattern between NSE parameters and finite-nuclei observables, and to
provide uncertainty quantification on theoretical predictions
using the powerful methods of statistical analysis \cite{Rei10}.

A second aim is to understand the dependences from a formal
 perspective and to explore the impact of configuration mixing.
Within the independent particle picture the isovector  response can be 
described in terms of a charge-dependent
symmetry potential that shifts the neutron well with respect to the proton average potential.
The effect can be estimated quantitatively within the Fermi-gas model (FGM) augmented by
a schematic isospin-isospin interaction~\cite{(Boh69)}
\begin{equation}\label{vTT}
V_{TT}= \frac{1}{2} \kappa \hat{\vec T}\cdot \hat{\vec T}.
\end{equation}
In the Hartree approximation this model gives rise to a
quadratic dependence of the NSE on the neutron excess $I = (N-Z)/A$: 
\begin{equation}\label{fermigas}
E_{\rm sym}/A =  a_{\rm sym} I^2 = (a_{\rm sym,kin} + a_{\rm sym,int}) I^2,
\end{equation}
as  $T=|T_z|=|N-Z|/2$ in the ground-states of almost all nuclei. 
The FGM, in spite of its simplicity, has played an important  role in our understanding of the NSE. 
In particular, it separates the NSE strength into kinetic and interaction (potential)  contributions, and predicts 
a near-equality  $a_{\rm sym,kin} \approx a_{\rm sym,int}$ of
these contributions. It also provides an estimate $a_{\rm sym}\approx 25$\,MeV for the NSE coefficient (see Ref.~\cite{(Mek12)} for a recent discussion).

Furthermore, we note that the SCMF approach can lead
  to spontaneous breaking of symmetries. This apparent drawback can
  be turned into an advantage, as the symmetry breaking mechanism
allows to incorporate many inter-nucleon correlations within a single
product state or, alternatively, within a single-reference DFT sacrificing
good quantum numbers; broken symmetries have to be
restored {\em a posteriori}. We will address this topic using the example of
isospin mixing which naturally has an impact on isovector properties.

This survey is organized as follows. Section~\ref{sec:DFT} outlines the SCMF approaches and details various theoretical ingredients of the models employed in this work.
Observables pertaining to bulk NM and finite nuclei that are essential for NSE are discussed in Sec.~\ref{sec:observ}. Constraints on NSE and correlations between observables and NSE parameters, using the statistical covariance technique, are  presented in Sec.~\ref{sec:correl}. Section \ref{sec:symmenpar} summarizes the current status of NSE parameters. The planned extensions of the current DFT work are laid out in Sec.~\ref{sec:isocrank}. Finally, Sec.~\ref{sec:concl} contains the conclusions of this survey.

\section{Nuclear DFT}
\label{sec:DFT}

The nuclear EDF constitutes a crucial ingredient for a set of DFT-based
theoretical tools that enable an accurate description of
ground-state properties, collective excitations, and large-amplitude
dynamics over the entire chart of nuclides, from relatively light
systems to superheavy nuclei, and from the valley of $\beta$-stability
to the nucleon drip-lines.  In general EDFs are not directly related to any
specific microscopic inter-nucleon interaction,
but rather represent universal functionals of nucleon densities and
currents. With a small set of global parameters adjusted to empirical
properties of nucleonic matter and to selected data on finite nuclei \cite{(Klu09),(Kor10)},
models based on EDFs enable a consistent description of a variety of
nuclear structure phenomena.

The unknown exact and  universal nuclear
EDF is approximated by simple, mostly analytical, functionals built from
powers and gradients of nucleonic densities and currents,
representing distributions of matter, spins, momentum and kinetic
energy. When pairing correlations are included, they are represented by pair (anomalous) densities.  
In the field of nuclear structure this method is analogous to Kohn-Sham DFT. SCMF models {\em effectively} map the nuclear many-body problem onto a one-body problem using auxilliary Kohn-Sham single-particle orbitals. By including many-body correlations in EDF, the Kohn-Sham method in
principle goes beyond the Hartree-Fock (HF) or Hartree-Fock-Bogolyubov (HFB) approximations and, in addition,
it has the advantage of using {\em local} potentials. A broad range of
nuclear properties have been very successfully described using SCMF
models based on Skyrme EDFs, relativistic EDFs, and the Gogny interaction
\cite{(Ben03),Sto07aR,(Erl11),(Lal04),(Vre05),(Men06),(Nik11)}. 
(Note that the Gogny model is not strictly local as the other EDFs.)
In the remainder of this section we briefly outline the Skyrme-Hartree-Fock (SHF) method and the
relativistic mean-field (RMF) approach. As both methods are widely used
and extensively described in the literature, we keep the presentation 
short and concentrate on a side-by-side comparison of the models. 

The basis of any mean-field approach is a set of single-nucleon
canonical (Kohn-Sham) orbitals $\psi_\alpha(\mathbf{r})$, with occupations amplitudes
$v_\alpha$. The $\psi_\alpha$ denote Dirac four-spinor wave functions in
the RMF framework, and two-component-spinor wave functions in the SHF
which is a classical mean-field model. The canonical occupation amplitudes
$v_\alpha$ are determined by the  pairing interaction. 
The starting point of a particular model is an EDF expressed in terms of
$\psi_\alpha,v_\alpha$ and the local densities derived therefrom.
The energy functional for the SHF method reads
\begin{eqnarray}
  E
  &=&
  \!\int\! d^3r\,\left(
    {\mathcal E}_{\rm kin}
    +{\mathcal E}_{\rm pot}
  \right)
  +
  E_{\rm Coul}
  +
  E_{\rm pair}
  +
  E_{\rm cm},
\label{eq:Skenfun}
\\
  {\mathcal E}_{\rm kin}
  &=& 
  \frac{\hbar^2}{2m_\mathrm{p}}\tau_\mathrm{p}
    +
    \frac{\hbar^2}{2m_\mathrm{n}}\tau_\mathrm{n}
\nonumber\\
  E_{\rm cm}^{\mbox{}}
  &=&
  -\frac{1}{2mA}\langle\big(\hat{P}_\mathrm{cm}\big)^2\rangle.
\nonumber
\end{eqnarray}
The kinetic energy ${\mathcal E}_{\rm kin}$ is expressed in terms of single-nucleon
wave functions. The Skyrme functional is contained in the interaction part
with the potential-energy density $\mathcal{E}_{\rm pot}$.  The
Coulomb energy $ E_{\rm Coul}$ consists of the direct Coulomb term,  
and the Coulomb exchange that is usually taken into account at the level of the Slater
approximation. In most applications the center-of-mass correction $E_{\rm cm}^{\mbox{}}$ is applied 
{\em a posteriori} because its variation would considerably complicate the mean-field
equations. The pairing functional $E_{\rm pair}$ will be detailed
later. The RMF approach is usually formulated in terms of a Lagrangian:
\begin{align}
{L} &= 
  \!\int\! d^3r\,\left(
    \mathcal{L}_{\rm kin}
    -\mathcal{E}_{\rm pot}
  \right)
  -
  E_{\rm Coul}
  -
  E_{\rm pair}
  -
  E_{\rm cm},
\label{Lagrangian}
\\
  \mathcal{L}_{\rm kin}
  &=
  \sum_\alpha v_\alpha^2\psi^\dagger_\alpha\hat{\gamma}_0 
  (i\hat{\bm{\gamma}} \cdot \bm{\partial} -m)\psi^{\mbox{}}_\alpha,
\end{align}
where $\hat{\gamma}$ is the Dirac
matrix. Again, the kinetic part is expressed explicitly in terms
of Dirac spinor wave functions, whereas interaction terms are included in the
potential energy density $\mathcal{E}_{\rm pot}$. Further
contributions from Coulomb, pairing and center-of-mass motion are
treated similarly as in the SHF approach.

The basic building blocks of an EDF are local densities and currents
built from single-nucleon wave functions \cite{Eng75a,(Ben03)}. These are summarized in
the upper part of Table~\ref{tab:sum-models}.
\begin{table*}
\begin{center}
\begin{tabular}{|cc|cc|}
\hline 
\multicolumn{4}{|c|}{\bf densities}
\\
\hline
\multicolumn{2}{|c}{SHF} &\multicolumn{2}{|c|}{RMF} \\ 
$T=0$ & $T=1$ & $T=0$ & $T=1$ \\ 
\hline 
$\displaystyle\rho_0({\bf r}) =\sum_{\alpha}v_\alpha^2\,
  \psi_\alpha^\dagger\psi_\alpha^{\mbox{}}$ 
& 
$\displaystyle\rho_1({\bf r}) =\sum_{\alpha}v_\alpha^2\,
\psi_\alpha^\dagger\hat{\tau}_3\psi_\alpha^{\mbox{}}$
&
$\displaystyle\rho_0({\bf r}) =\sum_{\alpha}v_\alpha^2\,
  \psi_\alpha^\dagger\psi_\alpha^{\mbox{}}$ 
& 
$\displaystyle\rho_1({\bf r}) =\sum_{\alpha}v_\alpha^2\,
\psi_\alpha^\dagger\hat{\tau}_3\psi_\alpha^{\mbox{}}$
\\
$\displaystyle\tau_0({\bf r}) =\sum_{\alpha}v_\alpha^2\,
  {\bm \nabla}\psi_\alpha^\dagger{\bm \nabla}\psi_\alpha^{\mbox{}}$ 
& 
$\displaystyle\tau_1({\bf r}) =\sum_{\alpha}v_\alpha^2\,
{\bm \nabla}\psi_\alpha^\dagger\hat{\tau}_3
{\bm \nabla}\psi_\alpha^{\mbox{}}$
&
$\displaystyle\rho_\mathrm{S}({\bf r}) =\sum_{\alpha}v_\alpha^2\,
  \psi_\alpha^\dagger\hat{\gamma}_0\psi_\alpha^{\mbox{}}$ 
&
\\
$\displaystyle\mathbf{J}_0({\bf r})=
  -\mathrm{i}\sum_{\alpha}v_\alpha^2\,
  \psi_{\alpha}^\dagger{\bm \nabla}\!\times\!\hat{\bsigma}
  \psi_{\alpha}^{\mbox{}}$
&
$\displaystyle\mathbf{J}_1({\bf r})=
  -\mathrm{i}\sum_{\alpha}v_\alpha^2\,
  \psi_{\alpha}^\dagger\hat{\tau}_3
 {\bm \nabla}\!\times\!\hat{\bsigma}\psi_{\alpha}^{\mbox{}}$
&&\\[8pt]
\hline
\end{tabular}
\\[6pt]
\begin{tabular}{|l|cc|cc|cc|}
\hline 
\multicolumn{7}{|c|}{\bf potential-energy density}
\\
\hline
 &\multicolumn{2}{|c}{SHF} &\multicolumn{2}{|c}{RMF-PC}
  &\multicolumn{2}{|c|}{RMF-ME} \\
 & $T=0$ & $T=1$ & $T=0$ & $T=1$ & $T=0$ & $T=1$ \\ 
\hline
$\rho\rho$ &  $C_0^\rho\rho_0^2$    & $C_1^\rho\rho_1^2$ 
  & $G_\omega\rho_0^2$ & $G_\rho\rho_1^2$ 
  & $G_\omega\rho_0\frac{1}{-\Delta+m_\omega^2}\rho_0$
  & $G_\rho{\rho_1} \frac{1}{-\Delta+m_\rho^2}{\rho_1}$
\\
mass &  $C_0^\tau\rho_0\tau_0$  & $C_1^\tau{\rho_1}{\tau_1}$ 
  & $G_\sigma\rho_\mathrm{S}^2$ &
  & $G_\sigma\rho_\mathrm{S}\frac{1}{-\Delta+m_\sigma^2}\rho_\mathrm{S}$ &
\\
${\bm\ell}\cdot{\bm s}$ & $C_0^{{\bm \nabla}\mathbf{J}}\rho_0{\bm \nabla}\!\cdot\!\mathbf{J}_0$ &
      $C_1^{{\bm \nabla}\mathbf{J}}{\rho}_1{\bm \nabla}\!\cdot\!{\mathbf{J}_1}$  
  & `` & & `` & \\
gradient & $C_0^{\Delta\rho}({\bm \nabla}\rho_0)^2$
  & $C_1^{\Delta\rho}({\bm \nabla}{\rho_1})^2$
  & $f_\mathrm{S}({\bm \nabla}\rho_\mathrm{S})^2$ & &&
\\
\hline
dens.dep. & $C_0^\rho=c_0^\rho+d_0^\rho\rho_0^a$ 
& $C_1^\rho=c_1^\rho+d_1^\rho\rho_0^a$
&\multicolumn{2}{|c|}{
 $\displaystyle G_i=a_i +(b_i+c_ix)e^{-d_ix}$}
  &  $\displaystyle G_i=a_i\frac{1+b_i(x+d_i)^2}{1+c_i(x+d_i)^2}$
  &  $\displaystyle G_\rho=g_\rho e^{-a_\rho(x-1)}$ 
\\
&\multicolumn{2}{|c|}{$C_T^\tau=c_T^\tau\;,\;
    C_T^{\Delta\rho}=c_T^{\Delta\rho}\;,\;
   C_T^{{\bm \nabla}\mathbf{J}}=c_T^{{\bm \nabla}\mathbf{J}}$}
 &\multicolumn{2}{|c|}{ $i\in\{\sigma,\omega,\rho\}\,$,
  $\,\displaystyle x=\frac{\rho_0}{\rho_\mathrm{sat}}$}
  &  $i\in\{\sigma,\omega\}\,$, 
    $\,\displaystyle x=\frac{\rho_0}{\rho_\mathrm{sat}}$
  &
\\[8pt]
\hline
\end{tabular}
\end{center}
\caption{\label{tab:sum-models}
Upper: The basic isoscalar ($T=0$) and isovector ($T=1$)  local densities of SHF (left) and RMF (right).
Lower:
The potential-energy densities in the three considered
SCMF models. Model parameters (third row) defining the coupling constants are indicated by lowercase latin letters.
For further explanation see text.
}
\end{table*}
All densities appear in two flavors \cite{(Per04),(Roh10)}: isoscalar ($T=0$), or total density (sum of proton 
and neutron densities), and isovector ($T=1$)  density (difference between neutron and proton densities). 
Both can be conveniently 
expressed using the isospin operator $\hat{\tau}_3$.  The
basic ingredients of an EDF are the local densities $\rho_0$ and $\rho_1$. 
In RMF these can be associated with the
zero-component of the four-vector current, where $\rho_0$ is often
called the vector density and $\rho_1$ the isovector-vector density.
RMF uses one more ingredient, the isoscalar-scalar density
denoted here as $\rho_\mathrm{S}$. SHF instead employs the kinetic-energy
densities $\tau_{0/1}$ and the spin-orbit densities
$\mathbf{J}_{0/1}$. One can show that $\tau_0$ and $\mathbf{J}_{0}$
emerge in the non-relativistic limit of $\rho_\mathrm{S}$
\cite{Rei89aR}. The principal difference between SHF and RMF is that 
the quantities $\tau_0$ and $\mathbf{J}_{0}$
are independent in SHF, whereas they are tightly related through
$\rho_\mathrm{S}$ in RMF. Moreover, the RMF does not invoke an
isovector counterpart of $\rho_\mathrm{S}$ thus being more restricted
in the isovector channel.

The lower part of Table~\ref{tab:sum-models} displays 
the main components of the potential-energy density. The underlying   is to take all
bi-linear isoscalar combinations of the local densities and to
associate a coupling constant with each term \cite{(Roh10)}. The SHF confines the
combinations to have at most second order of derivatives (the term
$\mathbf{J}^2$ is also dropped). In the RMF approach one keeps only terms that form a
Lorentz scalar. Moreover, two bi-linear realizations of RMF will be considered.
First there is the straightforward point-coupling (RMF-PC) realization 
that corresponds to contact interactions between nucleons and, second, 
the meson-exchange folding (RMF-ME). The folding is 
motivated by the traditional route to RMF as a
model of nucleons coupled to classical meson fields.  Of course, 
at energies characteristic for nuclear binding meson exchange represents 
just a convenient representation of the effective nuclear interaction. In 
practice RMF-PC and RMF-ME present equivalent realizations of the 
relativistic SCMF, differing in the range of effective interactions 
(zero-range vs. finite-range) and the choice of density dependence 
for the couplings. In practical applications one restricts the density dependence 
of coupling (vertex) functions 
to keep the number of free parameters to a minimum. In SHF, only the
leading terms $\propto\rho_0^2$ and $\rho_1^2$ are given a (simple)
density dependence as shown in Table~\ref{tab:sum-models}. In RMF-PC
and RMF-ME, each term has some density dependence, but not all of these
parameters are actually used. In RMF-PC, in particular, 
$c_1,a_\mathrm{S}$ and $c_\mathrm{S}$ are set to zero \cite{(Nik08)}. In RMF-ME,
the parameters are correlated by additional
boundary conditions on $G_i$ \cite{(Nik02),(Lal05)}. In total,
there are 11 adjustable parameters for SHF, 10 for RMF-PC, and 8
for RMF-ME. From a formal perspective, SHF and RMF-PC are rather similar, differing
mainly in the relativistic kinematics, while RMF-ME includes a significantly different density dependence 
of the couplings, in addition to the finite range. These three models thus allow to display separately effects of kinematics, density dependence, and range of the effective nuclear interaction.

As far as particle-particle interaction, in the SHF we use the pairing functional derived from a density-dependent zero-range force:
\begin{subequations}
\begin{eqnarray}
  E_{\rm pair}
  &=&
  \frac{1}{4} \sum_{q\in\{p,n\}}\int d^3r \tilde{\rho}^2_q
  \left[1 -\frac{\rho(\mathbf{r})}{\rho_\mathrm{pair}}\right],
\\
  \tilde{\rho}_q({\bf r})
  &=&
  \sum_{\alpha\in q} u_\alpha v_\alpha
  \big|\psi_\alpha({\bf r})\big|^2,
\end{eqnarray}
\end{subequations}
where $q$ runs over over protons and neutrons. It involves the
pair-density $\tilde{\rho}_q$ and is usually augmented by some density
dependence. We consider here $v_{0,p}$, $v_{0,n}$, and
$\rho_\mathrm{pair}$ as free parameters of the pairing functional
in SHF. Note that we do not recouple to isoscalar and isovector terms
because pairing is considered independently for protons and neutrons.
Actually, the zero-range pairing force works only together with a
limited phase space for pairing. We use here a soft cut-off 
in the space of single-nucleon energies \cite{Bon85a} according to Ref.~\cite{Ben00a}.

In RMF calculations we use the recently developed 
separable pairing force \cite{(Tia09a),(Tia09b)}. 
It is separable in momentum
space, and is completely determined by two parameters that are adjusted to
reproduce in symmetric nuclear matter the pairing gap
of the Gogny force. We have verified that both pairing prescriptions yield
comparable results for the pairing gaps.

\section{Observables}
\label{sec:observ}

In this section, we discuss  observables pertaining to nuclear matter (NM) and finite nuclei that are essential for discussion of NSE. Those observables can be roughly divided \cite{Rei10} into 
good isovector
indicators  that correlate very well with NSE (such as weak-charge form factor,
neutron skins, dipole polarizability,  slope of the symmetry energy, and neutron  pressure) and poor isovector indicators (such as nuclear and neutron matter binding energy,  giant resonance energies, isoscalar and isovector
effective mass, incompressibility, and saturation density).

\subsection{Nuclear matter properties}
\label{sec:symmenerg}

Bulk properties of symmetric nuclear matter, called nuclear matter
properties (NMP), are often used to characterize the properties of a
model, or functional respectively. Starting point for the definition
of NMP is the binding energy per nucleon in the symmetric nuclear matter
$E/A=E/A(\rho_0,\rho_1,\tau_0,\tau_1)$.  
\begin{table}
\begin{center}
\begin{tabular}{lrcl}
 incompressibility:
 &
  $  K_\infty$
  &=& 
  $\displaystyle
  9\,\rho_0^2 \, \frac{d^2}{d\rho_0^2} \,
       \frac{{E}}{A}\Big|_\mathrm{eq}
  $
\\[12pt]
  symmetry energy:
  &
  $a_\mathrm{sym}$
  &=&
  $\displaystyle
 \frac{1}{2} \frac{d^2}{d\rho_1^2}
  \frac{{E}}{A} \bigg|_\mathrm{eq}
  $
\\[12pt]
  slope of $a_\mathrm{sym}$:
  &
  $L$
  &=&
  $\displaystyle
  3\rho_0 \frac{d a_\mathrm{sym}}{d\rho_0} \bigg|_\mathrm{eq}
  $
\\[12pt]
  effective mass:
  &
  $\displaystyle
  \frac{\hbar^2}{2m^*}
  $
  &=&
  $\displaystyle
   \frac{\hbar^2}{2m}
    + 
    \frac{\partial}{\partial\tau_0} \frac{{E}}{A}\bigg|_\mathrm{eq}
  $
\\[12pt]
  TRK sum-rule enhanc.:
  &
  $\kappa_\mathrm{TRK}$
  &=& 
  $\displaystyle
  \frac{2m}{\hbar^2}
  \frac{\partial}{\partial\tau_1} 
  \frac{{E}}{A}\bigg|_\mathrm{eq}
  $
\end{tabular}
\end{center}
\caption{\label{tab:nucmatdef}
Definitions of NMP used in this work.
All derivatives are to be  taken at the equilibrium point corresponding to the saturation density $\rho_\mathrm{eq}$.
}
\end{table}
Table \ref{tab:nucmatdef} lists the NMP discussed in this work. It is important to note
the difference between total derivatives used for $K_\infty$,
$a_\mathrm{sym}$, $L$, and partial derivatives used for
$m^*/m$ and  $\kappa_\mathrm{TRK}$.  The latter take $E/A$ with $\tau_T$
as independent variables while the total derivatives employ the
dependence $\tau_T=\tau_T(\rho_0,\rho_1)$.
The slope of the symmetry energy $L$ parametrizes
the density dependence of $a_\mathrm{sym}$. This quantity is essential 
for the characterization  of the EOS of neutron matter and the mass-radius relation  
in neutron stars \cite{(Lat12),(Tsa12),(Fat12),(Fat12a),(Erl13),(Ste13)}.
The enhancement
factor $\kappa_\mathrm{TRK}$ for the Thomas-Reiche-Kuhn (TRK) sum
rule \cite{(Rin00)} characterizes the isovector effective mass.

Next to NMP come the corresponding bulk surface parameter, the
(isoscalar) surface energy $a_\mathrm{surf}$ and the (isovector)
surface-symmetry energy $a_\mathrm{ssym}$. These surface
parameters can be  determined from the leptodermous
expansion of the
liquid drop model (LDM) energy per nucleon, ${\cal E}_{\rm LDM}=E_{\rm LDM}/A$,
in terms of inverse
radius ($\propto A^{-1/3}$) and neutron excess $I$ \cite{(Rei06)}:
\begin{equation}
\begin{array}{rclclcl}
{\cal E}_{\rm LDM}(A,I) 
  &=& \displaystyle
  a_\mathrm{vol} 
  &+& \displaystyle
  a_\mathrm{surf}A^{-1/3}
  &+& \displaystyle
  a_\mathrm{curv}A^{-2/3} \medskip\\
  &&&+&
  a_\mathrm{sym}{I^2}
  &+& \displaystyle
  a_\mathrm{ssym} A^{-1/3}{I^2} \medskip\\
  &&&&
  &+&  \displaystyle
  a_\mathrm{sym}^{(2)}I^4.
\end{array}
\label{eq:ldm}
\end{equation}
The LDM energy ${\cal E}(A,I)$ is obtained from the DFT calculation by
subtracting the fluctuating shell correction energy. The general strategy behind this correction and leptodermous expansion is detailed in Refs.~\cite{(Rei06),(Kor12)}. In essence, we combine
 NM calculations ($A=\infty$) with (shell corrected)
DFT calculations for a huge set of spherical nuclei and extract
the surface parameters by a fit to the expansion (\ref{eq:ldm}). 
Alternatively and simpler,
one can compute the surface energy and surface-symmetry
energy thourgh a semi-classical approximation
(extended Thomas-Fermi) for the semi-infinite nuclear matter
\cite{Eif94a}.  In this survey, we shall apply both strategies, the
semi-classical approach whenever RMF is involved.

An important parameter characterizing the pure neutron matter is the neutron  pressure
\begin{equation}\label{npressure}
P(\rho_n) = \rho_n^2 \frac{d}{d\rho_n} \left({E \over A}\right)_n,
\end{equation}
a quantity that is proportional to  the slope of the
binding energy of neutron matter at a given neutron density
(derivative of neutron EOS). As discussed below, $P$ is excellent isovector indicator.

\subsection{Observables  from finite nuclei}
\label{sec:moreobs}

The total energy of a nucleus $E(Z,N)$ is the most basic observable
described by SCMF. It is also the most important ingredient for calibrating the
functional, see Sec.~\ref{sec:chi2}. We ofter consider binding energy differences. Of great
importance for stability analysis are separation energies and 
$Q_\alpha$ values.
Another energy observable, potentially useful in the context of NSE, is the indicator
\begin{eqnarray}\label{dVpn}
  \delta V_{pn}
   &=&
  -\frac{1}{4}\left[E(N,Z)-E(N-2,Z) \right. \nonumber \\
  &-& \left. E(N,Z-2)+E(N-2,Z-2)\right] 
\end{eqnarray}
involving the
double difference of binding energies \cite{(Zha89)}. Since $\delta V_{pn}$
approximates the mixed partial derivative of binding energy with respect to $N$ and $Z$, for nuclei with an appreciable neutron excess, the
average value of $\delta V_{pn}$  probes the symmetry energy term of LDM \cite{(Sto07pn)}:
$
\delta V_{pn}^{\rm LDM} \approx 2 \left(a_{\rm sym} + a_{\rm ssym} A^{-1/3} \right) / A.$
That is, the shell-averaged trend of $\delta V_{pn}$  is determined by the symmetry and surface symmetry energy coefficients.

It has been shown in \cite{Rei92c} that effective SCMF provide a
pertinent description of the form factors in the momentum regime
$q<2q_\mathrm{F}$ where $q_\mathrm{F}$ is the Fermi momentum.  The key
features of the nuclear density are related to this low-$q$
range. The basic
 parameters characterizing nuclear density distributions are:  r.m.s. charge radius $r_\mathrm{C}$,
diffraction radius $R_\mathrm{C}$, and surface thickness
$\sigma_\mathrm{C}$ \cite{Fri82a}.  The diffraction radius $R_{\rm
C}$, also called the box-equivalent radius, parametrizes the gross
diffraction pattern which resemble those of a hard sphere of radius
$R_\mathrm{C}$ \cite{Fri82a}. The actual charge form factor  $F_\mathrm{C} (q)$ falls off
faster than the box-equivalent form factor $F_\mathrm{box}$. This is due to the finite
surface thickness $\sigma$ which, in turn, can be determined by
comparing the height of the first maximum of $F_\mathrm{box}$
with $F_\mathrm{C}$ from the realistic charge distribution.
The charge halo parameter $h_\mathrm{C}$ is composed from the three
basic charge form parameters and serves as a nuclear halo parameter
found to be a relevant measure of the outer surface
diffuseness \cite{Miz00a}.

The charge distribution is basically a measure of the the proton
distribution. It is only recently that the parity-violating electron scattering experiment PREX has provided some
information on the weak-charge formfactor $F_W(q)$ of $^{208}$Pb 
\cite{(Abr12),(Hor12)}. These unique data gives access to neutron
properties, such as the neutron r.m.s. radius
$r_\mathrm{n}$. Closely related and particularly sensitive to the
asymmetry energy is the neutron skin $r_\mathrm{skin}=
r_\mathrm{n}-r_\mathrm{p}$, which is the
difference of neutron and proton r.m.s. radii. 
(As discussed in Ref.~\cite{Miz00a}, it is better to define the neutron skin 
through neutron and proton diffraction radii and surface thickness. However, for well-bound nuclei, which do not exhibit halo features, the above definition of $r_\mathrm{skin}$ is
practically equivalent.)
Neutron radii and skins are excellent isovector indicators 
\cite{(Ton84),(Rei99),(Fur02),(Yos04),(Che05),(War09),Rei10,(Pie12),(Fat12)}
that help
to check and improve isovector properties of the nuclear EDF \cite{Rei10}.

Nuclear excitations are characterized by the strength
distributions $S_{JT}(E)$ where $J$ is the angular momentum of the
excitation, $T$ its isospin, and $E$ the excitation energy.  For
example, the cross section for photo-absorption is proportional to
$S_{11}(E)$. The strengths functions can be obtained  from
the excitation spectrum:
\begin{eqnarray}
  S_{JT}(E)
  &=&
  \sum_n E_n B_n(EJT)\delta_\Delta(E-E_n),
\end{eqnarray}
where $E_n$ is the excitation energy of state $n$, $B_n(EJT)$ the
corresponding transition matrix element of multipolarity $J$ and
isospin $T$, and $\delta_\Delta$ as finite width folding function -- if $S_{JT}(E)$ is calculated theoretically using, e.g., the random phase approximation (RPA).  In our RPA estimates, we
use an energy dependent width
$\Delta=\mbox{max}(\Delta_\mathrm{min},(E_n-E_\mathrm{thr})/E_\mathrm{slope})$
which simulates the broadening mechanisms beyond RPA. The parameters
for $^{208}$Pb are $\Delta_\mathrm{min}=0.2$ MeV, $E_\mathrm{thr}=10$
MeV, and $E_\mathrm{slope}=5$ MeV. The resulting spectral distributions for
heavy nuclei, as  $^{208}$Pb,  show one clear giant resonance peak
at $E_{\rm GR}(JT)$ for $(J,T)=(0,0), (1,1), (2,0)$.
We will consider these resonance energies as characteristic observables
of dynamical response in heavy nuclei. The strength functions $S_{JT}(E)$ in light nuclei
are much more fragmented and cannot be reduced to one single
characteristic number.

There are other key observables that can be extracted from the strength
distributions, in particular for the dipole case $S_{11}(E)$, namely
the electric dipole polarizability
\begin{equation}
  \alpha_\mathrm{D}
  =
  \sum_n E_n^{-1} B_n(E11)
  \quad,  
\end{equation}
and the TRK sum rule
\begin{equation}
  \sum_n E_n  B_n(E11)
  =
  \frac{\hbar^2}{2m}\frac{NZ}{A}\left(1+\kappa_\mathrm{TRK}(Z,N)\right),  
\end{equation}
which defines the sum-rule enhancement $\kappa_\mathrm{TRK}(Z,N)$. Note
that the latter is an observable in a specific finite nucleus and
differs somewhat from  $\kappa_\mathrm{TRK}$ in nuclear matter. In the following, we will consider
$\alpha_\mathrm{D}$ and $\kappa_\mathrm{TRK}$ for
$^{208}$Pb. In particular, it has been demonstrated 
\cite{Rei10,(Pie12)} that
$\alpha_\mathrm{D}$ strongly correlates with NSE; hence, it can serve as excellent isovector indicator thant can be precisely extracted from 
measured E1 strength \cite{(Tam11)}. On the other hand, the low-energy E1 strength, sometimes referred to as the pygmy dipole strength, exhibits weak collectivity. The
correlation between the accumulated low-energy strength and the symmetry energy is weak, and 
depends on the energy cutoff assumed \cite{Rei10,(Dao12),Rei13}.

Giant resonances are small amplitude excitations and belong to the
regime of linear response. The low energy branch of isoscalar
quadrupole excitations is often associated with large amplitude
collective motion along nuclear shapes with substantial quadrupole
deformation. Of particular importance is  nuclear fission, which determines existence of heavy and superheavy nuclei.
As a simple and  robust measure of fission, we shall
consider the axial fission barrier height in $^{266}$Hs.
Unlike actinides,  most superheavy nuclei have one single
fission barrier \cite{Erl12a,(Sta13),(War12)}, which simplifies the analysis for our puroposes. It has to be kept in mind that the inner barrier is
often  lowered by triaxial shapes, but this is not important for the study of large-amplitude nuclear deformability.

\section{Symmetry energy: constraints and correlations}
\label{sec:correl}

\subsection{Brief review of $\chi^2$ technique and correlation analysis}
\label{sec:chi2}

As discussed in Sec.~\ref{sec:DFT}, the nuclear EDF is characterized  by about  a dozen of coupling constants $\mathbf{p}=(p_1,...,p_F)$ that are determined by confronting DFT predictions with experiment. The standard procedure   is to adjust the parameters $\mathbf{p}$ to a large set of nuclear observables  in carefully selected
nuclei \cite{(Ben03),(Klu09),(Kor10),(Fat11),(Gao13)}. This is 
usually  done by the standard  least-squares optimization technique. Starting point is the  $\chi^2$ objective function
\begin{equation}\label{chi2}
\chi^2(\mathbf{p})
=
\sum_{\mathcal{O}}
\left(
\frac{\mathcal{O}^\mathrm{(th)}(\mathbf{p})
      -
      \mathcal{O}^\mathrm{(exp)}}
     {\Delta\mathcal{O}}
\right)^2,
\end{equation}
where ``th'' stands for the calculated values, ``exp'' for
experimental data, and $\Delta\mathcal{O}$ for adopted errors.  The
optimum parametrization $\mathbf{p}_0$ is the one which minimizes
$\chi^2$ with the minimum value $\chi^2_0=\chi^2(\mathbf{p}_0)$.
Around the minimum $\mathbf{p}_0$,
there is a range of ``reasonable'' parametrizations  $\mathbf{p}$ that can be
considered as delivering  a good fit, i.e.,   $\chi^2(\mathbf{p})\leq\chi^2_0+1$. As this range is usually
rather small, we can expand $\chi^2$ as
\begin{eqnarray}\label{chi2a}
  \chi^2(\mathbf{p})\!-\!\chi^2_\mathrm{0}
  &\approx&
  \sum_{i,j=1}^F (p_i\!-\!p_{i,0})\mathcal{M}_{ij}(p_j\!-\!p_{j,0}),
\\
  \mathcal{M}_{ij}
  &=&
  {\textstyle\frac{1}{2}}\partial_{p_i}\partial_{p_j}\chi^2|_{\mathbf{p}_0}.
\end{eqnarray}
The reasonable parametrizations thus fill the confidence ellipsoid
given by  
\begin{equation}\label{confidence}
  (\mathbf{p}-\mathbf{p}_0)\hat{\mathcal{M}}(\mathbf{p}-\mathbf{p}_0)
  \leq 1,
\end{equation}
see Sec.~9.8 of \cite{[Bra97a]}.
Given a set of parameters  $\mathbf{p}$, any observable
$A=\langle\hat{A}\rangle$ can be uniquely computed.  In this way, $A=A(\mathbf{p})$. The value $A$
thus varies within the confidence ellipsoid, and this results in some uncertainty  $\Delta A$.  Let us assume for simplicity that the
observable varies  weakly with $\mathbf{p}$ such that one can
linearize in the relevant range
$A(\mathbf{p})=A_0+(\mathbf{p}-\mathbf{p}_0)\cdot\bm{\partial}_\mathbf{p}A$.
Let us, furthermore, associate a weight
$\propto\exp{\left(-\chi^2(\mathbf{p})\right)}$ with each parameter set.
A weighted average over the parameter space yields the  covariance between two observables $\hat{A}$ and $\hat{B}$, which represents their
combined uncertainty:
\begin{equation}\label{cova}
  \overline{\Delta A\,\Delta B}
  =
  \sum_{ij}\partial_{p_i}A(\hat{\mathcal{M}}^{-1})_{ij}\partial_{p_j}B
\quad.
\end{equation}
For $A$=$B$, Eq.~(\ref{cova}) gives the variance $\overline{\Delta^2 A}$
that defines a statistical  uncertainty of an observable. 
Variance and covariance are useful quantities that allow to estimate
the impact of an observable on the model and its parametrization. 
We shall explore  the covariance analysis in
three different ways:
\begin{enumerate}
 \item We perform a constrained fit during which the observable of interest is
  kept fixed at a desired value. In the present survey, we consider the
  symmetry energy $a_\mathrm{sym}$ as constraining observable.  Comparing
  uncertainties from a constrained fit with those from an unconstrained fit
  provides a first indicator on the impact of the constrained observable on
 other observables.  
 \item 
  The next step is a trend analysis, in which one  performs a series of constrained fits
  with systematically varied values of the constraining observable. One then
  studies other observables as a function of the constrained quantity. This provides valuable information on possible inter-dependences.
\item
  Finally, we compute correlation (\ref{cova}) between
  $a_\mathrm{sym}$ and other observables. Here, a useful
  dimensionless measure is given by the Pearson product-moment correlation
  coefficient:
\cite{[Bra97a]}:
\begin{equation}
  {c}_{AB} 
  =
  \frac{|\overline{\Delta A\,\Delta B}|}
       {\sqrt{\overline{\Delta A^2}\;\overline{\Delta B^2}}}.
\label{correlator}
\end{equation}
  A value ${c}_{AB}=1$ means fully correlated and ${c}_{AB}=0$ --  uncorrelated.
\end{enumerate}
In the following, we will apply these three ways of studying correlations with 
 $a_\mathrm{sym}$ to  different groups of
observables. To this end, 
we have produced a series of parametrizations  with systematically varied
 $a_\mathrm{sym}$ for the SV Skyrme family
and for the  RMF-ME and RMF-PC models.

The optimization and covariance analysis carried out in this survey
is based for all three EDFs (SHF-SV, RMF-PC, and RMF-ME) on the same standard set of data on spherical nuclei (masses, diffraction radii, surface thickness, charge radii, separation energies, isotope shifts, and odd-even mass differences) that has originally been proposed in Ref.~\cite{(Klu09)} and recently employed in Refs.~\cite{(Erl10),(Erl13)}. We wish to emphasize that this is the first time that one consistent phenomenological input has been used to constrain SHF and RMF EDFs. A slightly modified  variant of the fitting
  protocol has been used for RMF-ME. This EDF did not lead to
  stable results in the fits which were unconstrained by NMP. Consequently, we
  included the nuclear matter information on $(E/A)_\mathrm{eq}$ into the
  dataset. This is still much less than in the previously published optimization protocols  of
  RMF-ME, in which  all NMP were constrained \cite{(Typ99),(Nik02),(Lal05)}.

\subsection{Correlations with nuclear matter properties}
\label{sec:nmp}

The NMP corresponding to unconstrained optimization of SHF-SV, RMF-PC, and RMF-ME EDFs -- using the same standard dataset -- are shown in Table~\ref{tableNMPFinal}. 
They are compared with NMP of SHF-RD \cite{(Erl10)} (employing a modified density dependence and the standard dataset) and SHF-TOV \cite{(Erl13)}
(using neutron star data in addition the standard dataset in the optimization process). 
As expected, isoscalar effective mass is significantly lowered in RMF as compared to SHF, and the opposite holds for $\kappa_{\rm TRK}$. The slope parameter $L$ is predicted to be very different in all five models. In particular, RMF-ME has very low value of $L$, and -- at the same time -- the uncertainty on $a_{\mathrm{sym}}$ in this model is very small.

\begin{table*}[ht]
\begin{center}
\begin{tabular}{|c|ccccccc|}
\hline
model & $\rho_{\rm eq}$ & $E/A$ & $K_\infty$ & $m^*/m$ &  $a_{\mathrm{sym}}$  & $L$ & $\kappa_{\rm TRK}$ \\[-5pt]
      & (fm$^{-3}$) & (MeV) & (MeV) &  & (MeV) & (MeV) & \\
\hline
SHF-SV & 0.161(1)  & -15.91(4) & 222(9) &  0.95(7)  & 31(2)  & 45(26) & 0.08(29)  \\
RMF-PC & 0.159(1)  & -16.14(3) & 185(18) & 0.57(1) & 35(2)  & 82(17) & 0.75(2) \\
RMF-ME & 0.159(3)  & -16.2(2) & 250(19) & 0.56(1) & 32.4(1)  & 6(7) & 0.79(2) 
\\\hline
SHF-RD & 0.161  & -15.93 & 231 &  0.90  & 32(2)  & 60(32) & 0.04(32)  \\
SHF-TOV & 0.161  & -15.93 & 222 &  0.94  & 32(1)  & 76(15) & 0.21(26) \\
\hline
\end{tabular}
\caption{\label{tableNMPFinal} Nuclear matter parameters of SHF-SV, RMF-PC, and RMF-ME EDFs used in this survey 
(with error bars) obtained by means of unconstrained optimization. Also shown are the values of NMP of SHF-RD \cite{(Erl10)} and SHF-TOV \cite{(Erl13)}. 
}
\end{center} 
\end{table*}

\begin{figure}
\centerline{\includegraphics[width=0.8\linewidth]{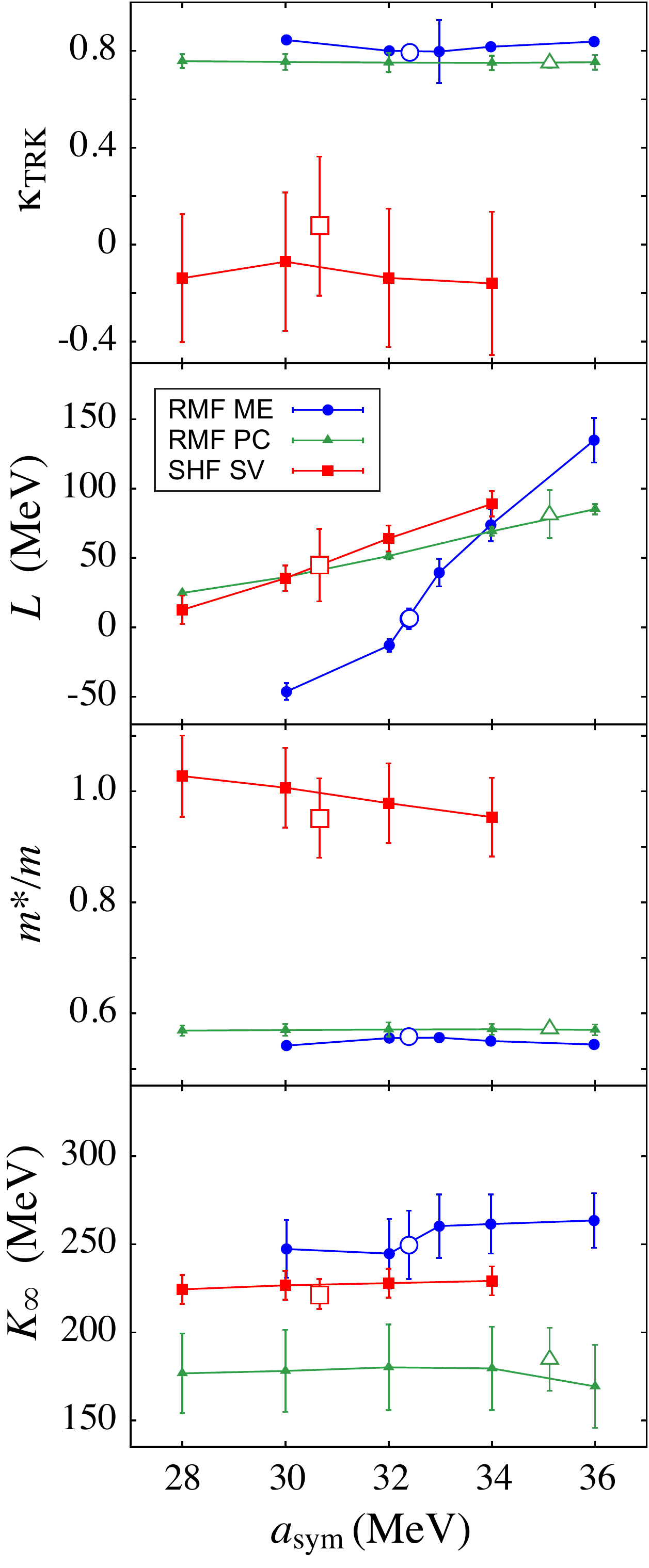}}
\caption{\label{fig:SV-symo-trends}
Behavior of selected nuclear matter properties with symmetry energy
$a_\mathrm{sym}$ for the SV Skyrme family
and for the ME and  PC RMF model families. 
The statistical uncertainties are
indicated by error bars. The result of the unconstrained fits are shown by 
large open symbols with corresponding error bars.
}
\end{figure}

Figure \ref{fig:SV-symo-trends} shows the trends for selected properties
of symmetric nuclear matter with $a_\mathrm{sym}$. 
The purpose of this analysis is to relate systematic variations with
$a_\mathrm{sym}$ to statistical 
uncertainties. The isoscalar properties $K_\infty$, $m^*/m$ as well as the isovector
dynamical response $\kappa_{\rm TRK}$ are fairly insensitive to
$a_\mathrm{sym}$. Their variation with $a_\mathrm{sym}$ are much smaller than
the  typical statistical uncertainties. This independency is also indicated by the fact that the uncertainty obtained in the unconstrained fit is not visibly larger than those from the constrained optimizations. The trend is markedly different for the density dependence of the symmetry energy $L$: 
variations with $a_\mathrm{sym}$  well exceed the statistical error bars  
and  the uncertainties from  unconstrained fits are larger
than those from constrained calculations.
It is to be noted that the dedicated variations of $a_\mathrm{sym}$  stay within the uncertainty of
$a_\mathrm{sym}$ in the unconstrained optimization. The uncertainty of $L$ in the free fit thus
covers nicely the uncertainty of the constrained calculations plus the variation of  $L$ with $a_\mathrm{sym}$. Anyway, the results shows that $L$ cannot be
  used as independent NMP although the formal structure of the EDF
  would allow that. There seems to be a strong link established by the data which yet has to be worked out.

\begin{figure}[htb]
\centerline{\includegraphics[width=0.7\linewidth]{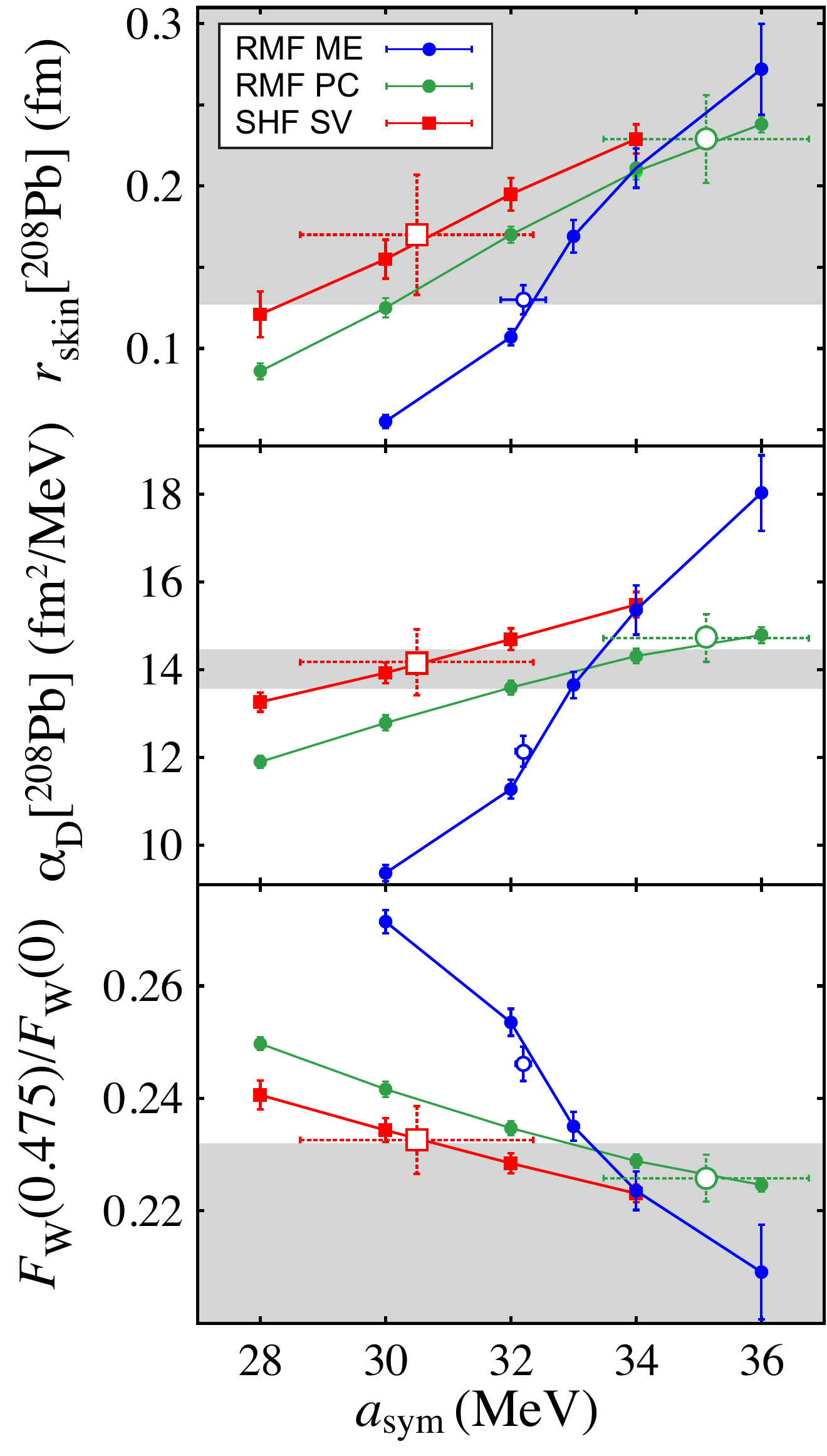}}
\caption{\label{fig:SV-symo-trendsfinprop} 
Similar as in Fig.~\ref{fig:SV-symo-trends} but for
selected   properties of $^{208}$Pb: neutron skin (top), dipole polarizability (middle), and weak-charge form factor (bottom). The current experimental ranges
are shaded  grey: $r_{\rm skin}$=0.33$^{+0.16}_{-0.18}$\,fm \cite{(Abr12)},
$\alpha_{\rm D}=14.0\pm 0.4$\,fm$^2$/MeV \cite{(Tam11)}, and 
$F_W(0.475)/F_W(0) = 0.204\pm 0.028$ \cite{(Hor12)}.
}
\end{figure}

\subsection{Correlations with properties of finite nuclei }
\label{sec:finnucl}

Figure \ref{fig:SV-symo-trendsfinprop} illustrates the trends with $a_\mathrm{sym}$
and extrapolation uncertainties for three observables in $^{208}$Pb:
weak-charge form factor at  $q= 0.475$\,fm$^{-1}$ ($q$-value of PREX),
neutron skin, and dipole polarizability.
These observables are all known to be
sensitive to isovector properties of EDF \cite{Rei10,(Pie12)}. This is confirmed by the
trends in the present result. The comparison of uncertainties shows a large
growth when going from  constrained to  unconstrained optimizations. This
corroborates the close relation between the symmetry energy and the three
isovectors indicators shown in Fig.~\ref{fig:SV-symo-trendsfinprop}. It is, furthermore, interesting to note that SHF and RMF-PC stay safely within the bands given by experimental data and RMF-ME is not far away. A better
discrimination between models requires more precise data, a task on which
presently many experimental groups are heavily engaged.

\begin{figure}[htb]
\centerline{\includegraphics[width=0.8\linewidth]{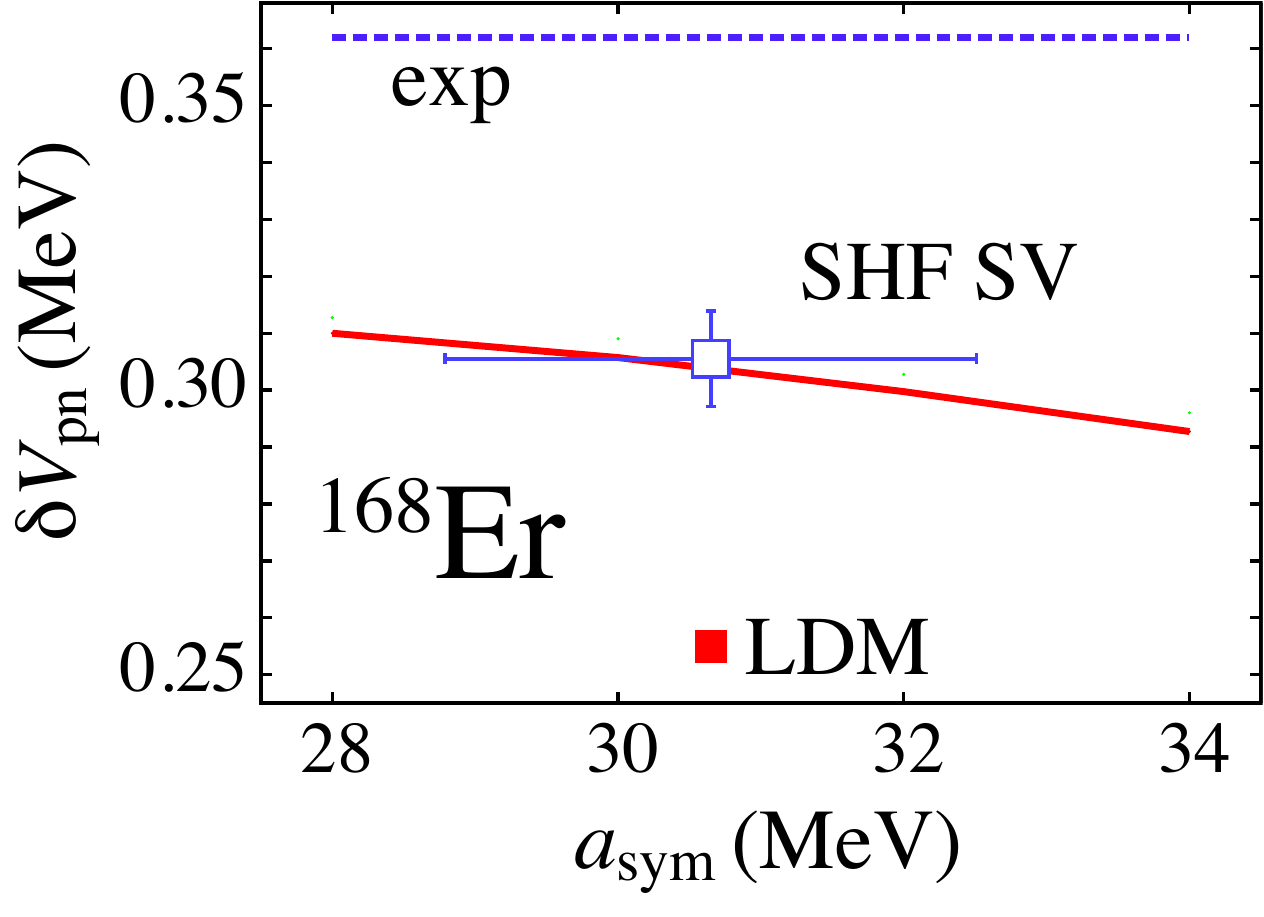}}
\caption{\label{fig:vpn} 
Behavior of $\delta V_{pn}$ in $^{168}$Er  with symmetry energy
$a_\mathrm{sym}$ for SHF-SV (solid line) as compared to experiment (dashed line) and the LDM value (filled square).
The result of the unconstrained fit is marked by 
a large open square with corresponding error bars. 
}
\end{figure}
To explore  the usefulness of $\delta V_{pn}$ as an isovector indicator, we choose the heavy deformed nucleus $^{168}$Er, as its even-even neighbors
have similar structure and the calculated values of $\delta V_{pn}$ for even-even Er isotopes show little variations around $N=100$. The results
 displayed in Fig.~\ref{fig:vpn} show a gradual decrease of this quantity with  $a_\mathrm{sym}$, but the magnitude of the variation is very small and cannot 
 account for the deviation from  experiment (around 50\,keV).  It is apparent that this quantity is too strongly influenced by shell effects (given by the deviation from the LDM estimate; also around 50\,keV) to probe NSE, see 
Refs.~\cite{(Sto07pn),(Ben11)} and Sec.~\ref{sec:correlsum} below.

Figure~\ref{fig:SV-symo-trendsGR} shows the trends of the three major
giant resonances in $^{208}$Pb: isoscalar monopole resonance (GMR),
isovector dipole resonance (GDR), and isoscalar quadrupole resonance
(GQR). For technical reasons, we only show results obtained with the SV Skyrme family.
\begin{figure}[htb]
\centerline{\includegraphics[width=0.8\linewidth]{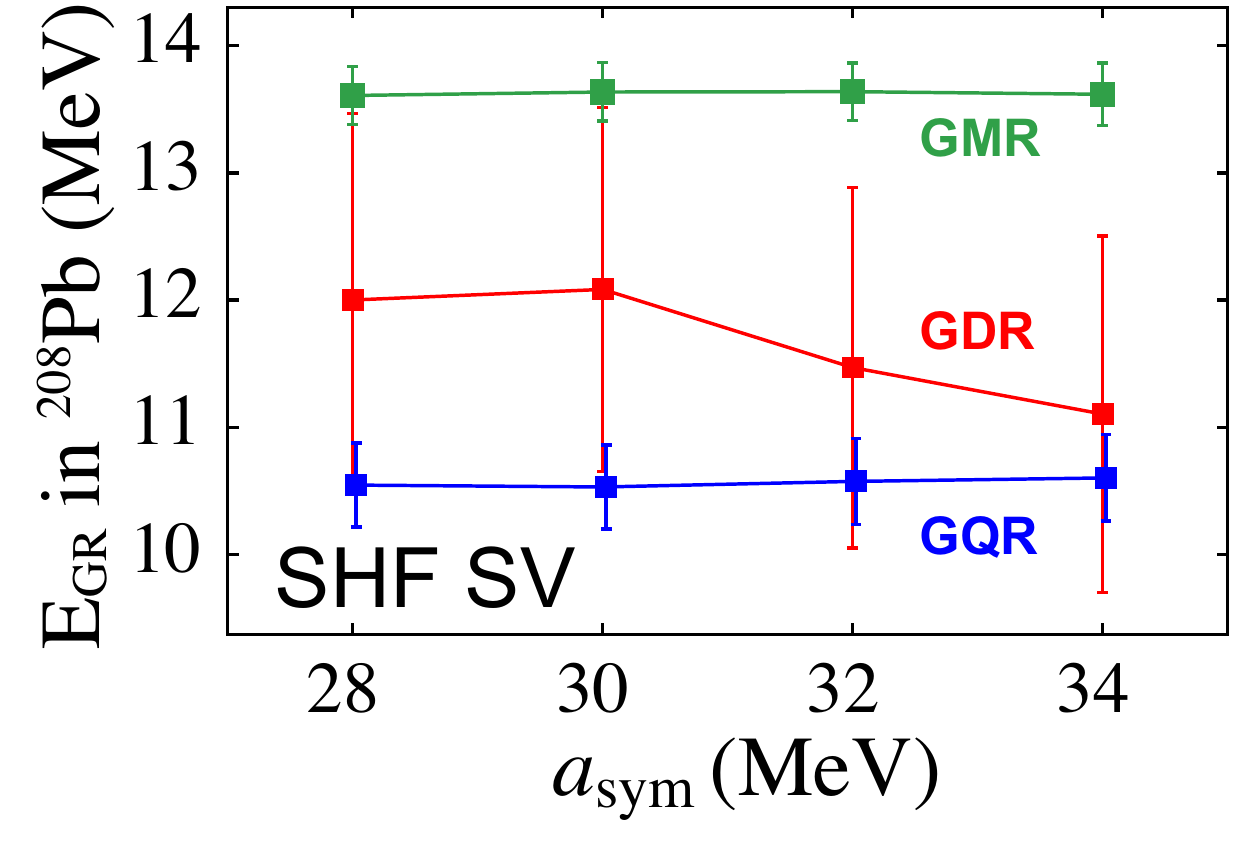}}
\caption{\label{fig:SV-symo-trendsGR} 
Behavior of giant resonance  energies in $^{208}$Pb  with symmetry energy
$a_\mathrm{sym}$ for the SV Skyrme family \cite{(Klu09)}.
In order not to make the graph too busy the uncertainties from the
unconstrained fit are not shown;  they have the same size as those from the
constrained fits.
}
\end{figure}
The isoscalar resonances show no dependence on $a_\mathrm{sym}$ at all; this is understandable for the symmetry energy belongs to the isovector
sector. Somewhat surprisingly, the GDR exhibits  very little
dependence on $a_\mathrm{sym}$ as well, with the magnitude of variations well  below the
statistical uncertainties. As demonstrated earlier \cite{(Klu09),(Pie12)}, it is 
the sum-rule
enhancement factor   $\kappa_\mathrm{TRK}$
that has 
the dominant impact on the GDR
peak frequency rather than $a_\mathrm{sym}$. The covariance analysis of Fig.~\ref{fig:SV-symo-trendsGR}
 confirms that the energies
of GMR, GDR, and GQR  do not obviously relate to $a_\mathrm{sym}$.

\begin{figure}
\centerline{\includegraphics[width=0.8\linewidth,angle=0]{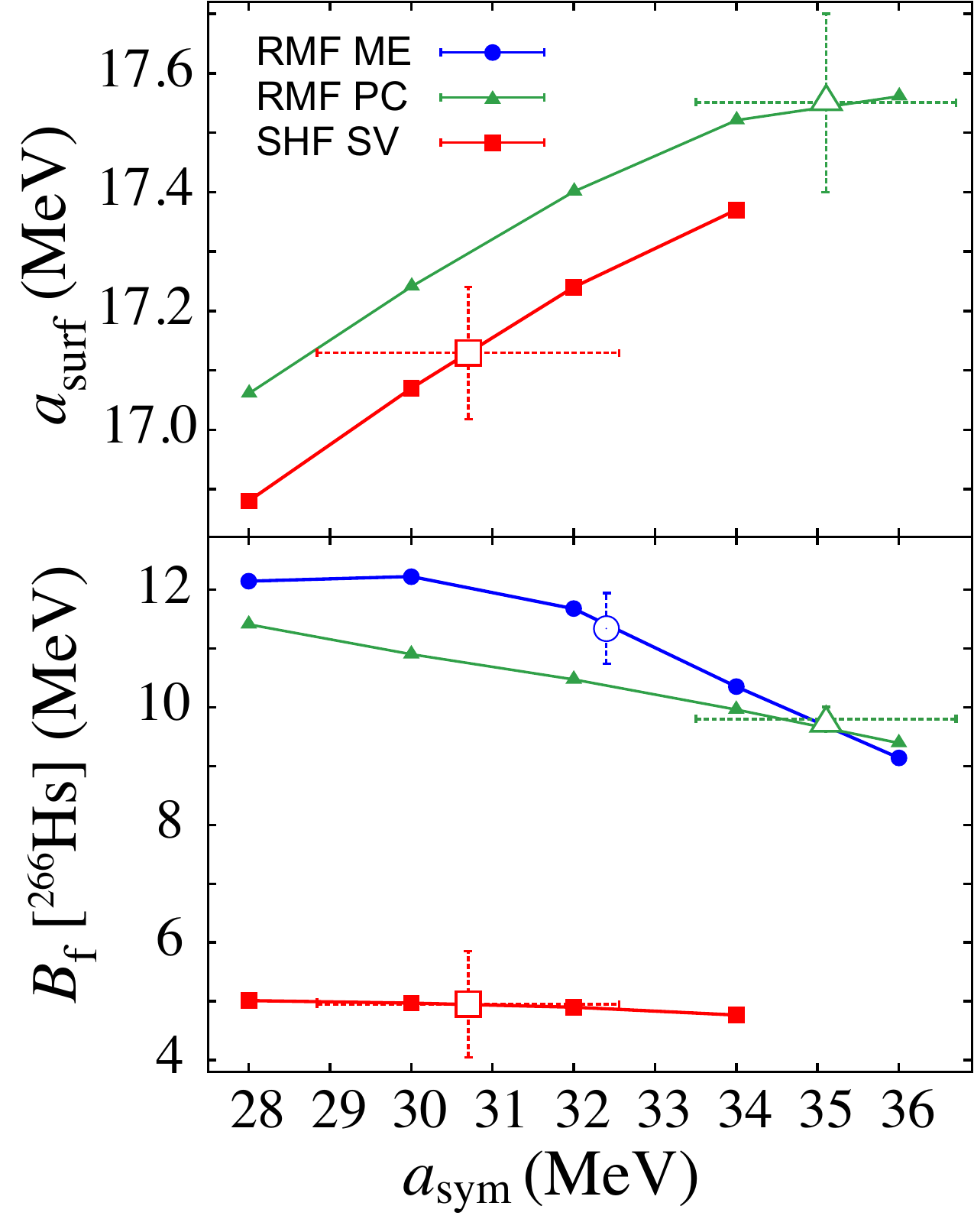}}
\caption{\label{fig:trends-barrier} 
Similar as in Fig.~\ref{fig:SV-symo-trends} but for
surface energy (top) and fission barrier
(bottom) in $^{266}$Hs. The surface energy from RMF-ME is not shown.
  }
\end{figure}
Figure \ref{fig:trends-barrier} shows behavior of surface
energy $a_\mathrm{surf}$ and the inner fission barrier $B_f$ in $^{266}$Hs with
$a_\mathrm{sym}$. The surface energy was
computed by means of the extended Thomas-Fermi method. The trends of $a_\mathrm{surf}$
 predicted by SHF and RMF are similar. An  offset of
about 2 MeV is most likely due to very different effective masses in both models.  Much
larger differences are seen for the fission barriers. The basic difference
between SHF and RMF can again be explained predominantly in terms of
effective masses. Barriers are produced by shell effects and shell
effects are larger for lower effective masses. There is also a
difference between the two RMF models. This could be due to a
different handling of gradient terms (only RMF-PC contains such) and a
much different parametrization of density dependence. All three models
show not only different values as such, but also different trends. 

The statistical errors  differ substantially between the models.  RMF-ME shows a  small uncertainty in $B_f$. This may be due to the missing gradient term in
this model which would also restrict the uncertainty in the surface
energy.  We note, however, that the gradient term in RMF-PC is to a certain extent 
equivalent the mass term of the sigma meson in RMF-ME, which is 
considered a free parameter. The plot of the $B_f$ demonstrates nicely the relative role of
statistical and systematic errors, with the statistical errors being
much smaller than inter-model differences. 
As discussed in Refs.~\cite{Nikolov11,(Kor12)}, fission barriers are strongly affected by $a_\mathrm{surf}$ and $a_\mathrm{ssym}$ of EDF. In particular,
the recently developed EDF UNEDF1, suitable for studies of strongly elongated nuclei, has relatively low values
of $a_\mathrm{surf}$ and $a_\mathrm{ssym}$ (see Fig.~\ref{assym} below) that  reflect the constraints on the fission isomer data.
The reduced surface energy coefficients result in a reduced effective 
surface coefficient
$
  a_{\rm surf}^{\rm (eff)} = 
 a_{\rm surf} + a_{\rm ssym} I^2,
$
which has profound consequences for the description of fission barriers, especially in the neutron-rich nuclei that are expected to play a role  
at the final stages of the r-process through the recycling mechanism \cite{(Pan08)}.

\subsection{Correlations summary}
\label{sec:correlsum}

\begin{figure}
\centerline{\includegraphics[width=0.8\linewidth]{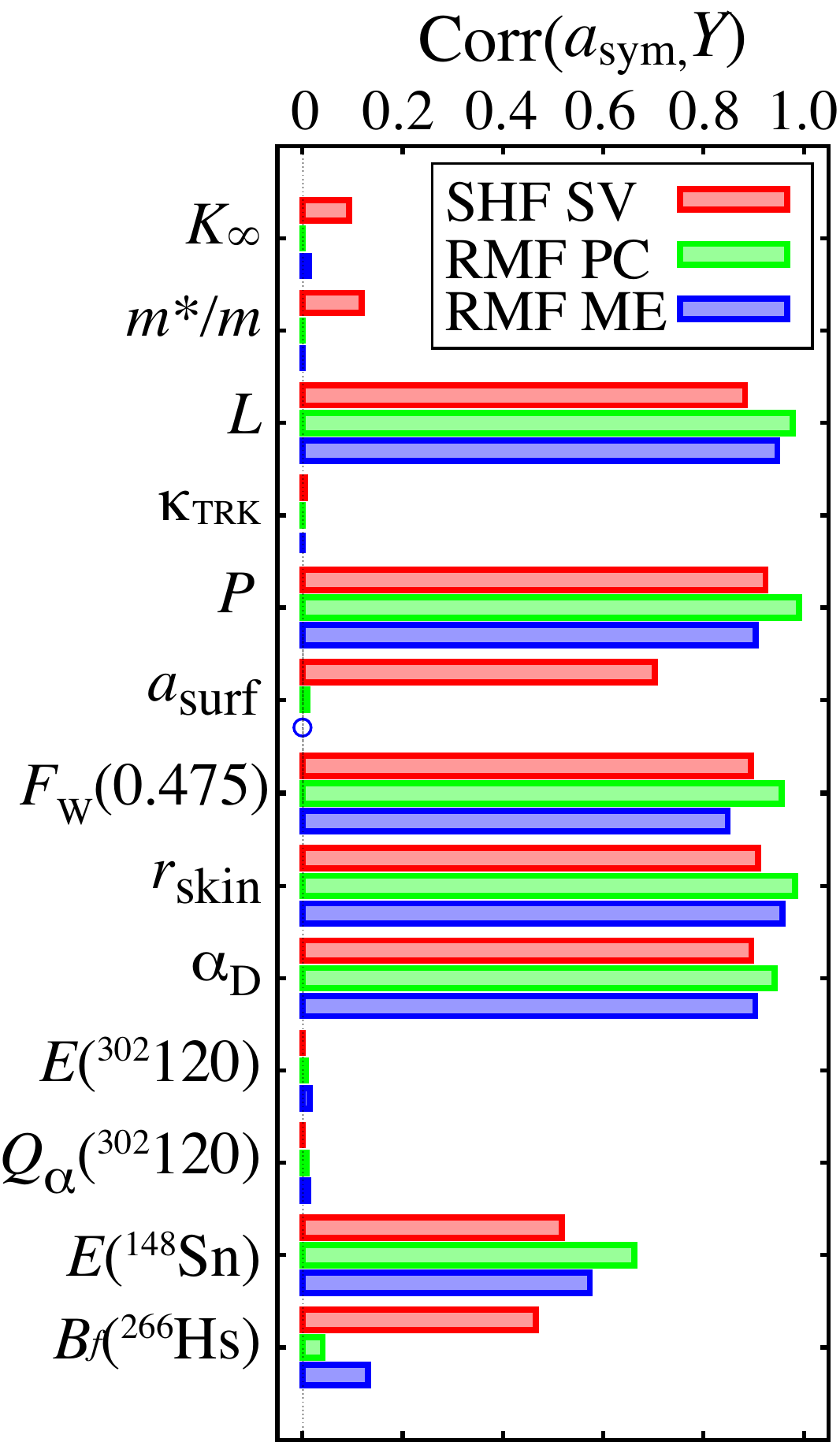}}
\caption{\label{fig:symo-correl} The correlation (\ref{correlator}) between symmetry energy  and selected observables ($Y$) for
  three models: SHF-SV, RMF-PC and RMF-ME.
  Results correspond to
  unconstrained optimization  employing  the same strategy in all three cases.
  For RMF-ME no reliable numbers could be obtained for $a_\mathrm{surf}$; this is indicated by an open circle.
  }
\end{figure}

The summary of our correlation analysis for $a_\mathrm{sym}$  is given in
Fig.~\ref{fig:symo-correl}.
The first four entries concern the same nuclear matter properties as in Fig.~\ref{fig:SV-symo-trends}.  It is only for $L$, the density dependence
of symmetry energy, that a strong correlation with $a_\mathrm{sym}$ is seen.  This complies nicely with the findings
of the trend analysis in Fig.~\ref{fig:SV-symo-trends}.
The next entry
concerns the neutron pressure (\ref{npressure})
at $\rho_n=0.08$\,neutrons/fm$^{3}$.  It is also strongly
correlated with $a_\mathrm{sym}$, which is no surprise because it is
an excellent isovector indicator \cite{(Bro00),(Typ01),(Fur02),(Yos04),Rei10,(Fat12)}. 
The diagram shows, furthermore, the (isoscalar) surface energy
$a_\mathrm{surf}$ computed in semi-classical approximation. This quantity is well correlated with $a_\mathrm{sym}$ for SHF and
practically uncorrelated for RMF. 

The next three entries are observables in $^{208}$Pb: weak-charge
form factor, neutron skin, and  dipole polarizability. All three are known to be strong isovector indicators \cite{Rei10,(Pie12),(Fat12)}. This is
confirmed here for all three models. 

The remaining four entries deal with exotic nuclei. These are:  binding energy and
$\alpha$-decay energy in yet-to-be-measured  superheavy nucleus $Z=120, N=182$,
binding energy in an extremely neutron rich $^{148}$Sn, and the
fission barrier in $^{266}$Hs (for which trends had been shown already
in Fig.~\ref{fig:trends-barrier}). The data on $Z=120, N=182$ consistently do not
correlate with $a_\mathrm{sym}$. The binding energy of
$^{148}$Sn shows some correlation with $a_\mathrm{sym}$, about equally
strong in the three models. This is expected as a large neutron
excess surely explores the static isovector sector.  Finally, the correlation with fission barrier in  $^{266}$Hs exhibits an appreciable  model dependence with some correlation in SHF and practically none in RMF.

We also studied correlations between $\delta V_{pn}$ in $^{168}$Er and other observables for finite nuclei and NM. We did not find a single observable that would correlate well with this binding-energy indicator. In particular, the correlation coefficient (\ref{correlator}) with $a_\mathrm{sym}$ is 0.41, with $\alpha_{\rm D}$ in $^{208}$Pb is 0.6, and with $r_{\rm skin}$ in $^{208}$Pb  is 0.54. This results demonstrates that $\delta V_{pn}$ in one single nucleus is too strongly influenced by shell effects to be used as an isovector indicator.

\section{Symmetry energy parameters of EDFs}
\label{sec:symmenpar}

The actual values of symmetry energy parameters depend on (i) the form of EDF and (ii) the optimization strategy used. The first point is nicely illustrated in Table~\ref{tableNMPFinal}, which compares NMP for different functional
forms (SHF-SV, SHF-RD, RMF-PC, and RMF-ME)
using the same dataset and
the same optimization technique. As far as the second point,  
it is instructive to compare SHF-SV and SHF-TOV NMP; namely, the inclusion of additional data on neutron stars in SHF-TOV has significantly impacted $L$ and
$\kappa_{\rm TRK}$. Many other examples can be found
in Refs.~\cite{Sto07aR,Dutra} that demonstrate divergent
predictions of Skyrme EDFs for neutron and nuclear matter.

The range of $a_{\text{sym}}$ is fairly  narrowly constrained by various data and ab-initio theory \cite{(Lat12)}; it is  $28 \,{\rm MeV}  <a_{\text{sym}}< 34$\,MeV.  The recent Finite-Range Droplet  Model (FRDM) result
\cite{(Mol12)} is  $a_{\text{sym}} = (32.5\pm 0.5)$\,MeV.
All EDFs listed in Table~\ref{tableNMPFinal} are consistent with these expectations. 

The values of $L$ are less precisely determined
\cite{(Li08),(Tri08),(Ste12),(Lat12),(Tsa12),(Fat12),(Fat12a),(Erl13),(Ste13),(Fat13)}; there is more dependence on specific observables or methodology  used.
Recent surveys \cite{(Lat12),(Ste13)} suggest that a reasonable range of $L$ is  $40\,{\rm MeV} < L< 80$\,MeV, and FRDM gives $L=70 \pm 15$\,MeV \cite{(Mol12)}.
Except for RMF-ME, all models shown in Table~\ref{tableNMPFinal} are consistent with these estimates. The low value of $L$ in RMF-ME is troublesome; here  we note that while
SHF-SV and RMF-PC EDFs fall within the error bars of the current experimental data  in Fig.~\ref{fig:SV-symo-trendsfinprop}, RMF-ME (as defined by the present  optimization protocol) does not.

As discussed in Ref.~\cite{(Rei06)}, the leading surface and symmetry terms appear 
relatively similar within each family of EDFs, with a clear difference for $a_{\mathrm{sym}}$ between  SHF and RMF.
By averaging over Skyrme-EDF results of Refs.~\cite{(Rei06),(Sat06)}, one obtains: $a_\mathrm{sym}\approx 30.9\pm 1.7$\,MeV,
$a_{\mathrm{ssym}} \approx -48\pm 10$\,MeV.
Older relativistic models provide systematically larger values~\cite{(Rei06)}:
 $a_\mathrm{sym}\approx 40.4\pm 2.7$\,MeV and 
$a_{\mathrm{ssym}} \approx -103\pm 18$\,MeV.
(Codes for a leptodermous expansion of the recent RMF-PC and RMF-ME models have yet to be developed.)

\begin{figure}
\centerline{\includegraphics[width=1.0\linewidth]{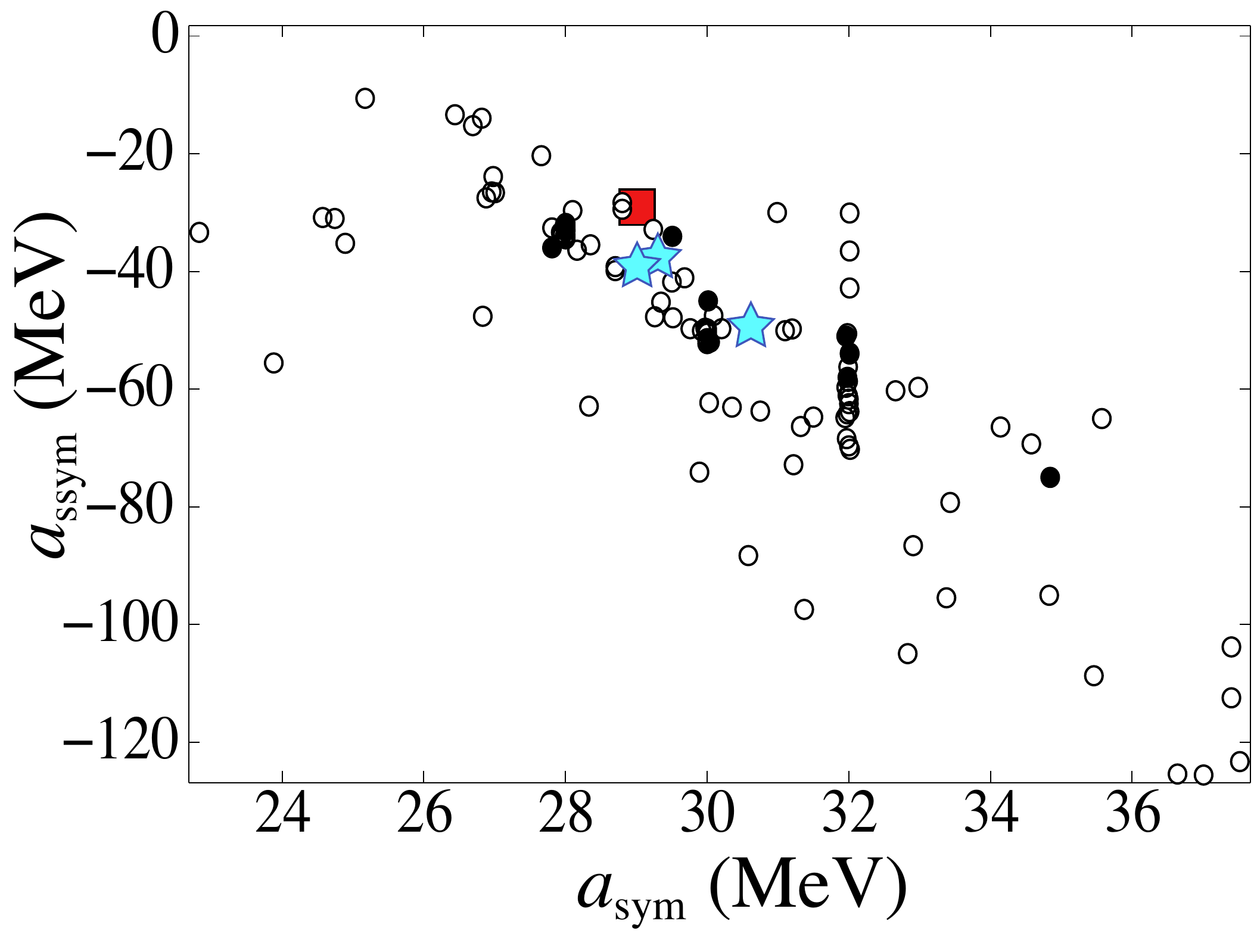}}
\caption{\label{assym} Correlation between the symmetry and
surface symmetry coefficients taken from
Ref.~\cite{Nikolov11} (Skyrme EDFs, dots; LDM values, stars) and Ref.~\cite{(Dan09)} (Skyrme EDFs, circles). The UNEDF1 values \cite{(Kor12)} are marked by a square. (Adopted from \cite{Nikolov11}.)
  }
\end{figure}

The coefficient $a_{\mathrm{ssym}}$ is  poorly constrained in the current EDF parameterizations and there are large differences between models, see Fig.~\ref{assym}.
In addition, the values of
 $a_\mathrm{sym}$ and  $a_\mathrm{ssym}$ have been shown to be systematically (anti)correlated  \cite{(Far78),(Ton84),(Dan09),Nikolov11}. 
Figure~\ref{assym}, 
displays the pairs $(a_{\text{sym}}, a_{\text{ssym}})$ for various  Skyrme EDFs 
and LDM parametrizations.  While a correlation between  $a_{\text{sym}}$ and
$a_{\text{ssym}}$ is apparent,
a very large spread of values is seen that demonstrates that the is indicative of the  data on 
g.s.  nuclear properties are not able to constrain $a_\mathrm{ssym}$.
It is interesting to note that the LDM values and phenomenological estimates cluster around $a_{\text{sym}}=30$\,MeV and
$a_{\text{ssym}}=-45$\,MeV. The values for UNEDF1 functional, additionally constrained 
by the data on very deformed fission isomers (thus probing the surface-isospin sector of EDF) are $a_{\text{sym}}=29$\,MeV and $a_{\text{ssym}}=-29$\,MeV.

\section{Isospin physics and symmetry energy}
\label{sec:outlook}

The emergence of NSE is rooted in the isobaric symmetry and its breaking
as a function of neutron excess and mass. Single-reference DFT is essentially the only framework allowing for understanding 
global behavior of  isospin effects throughout the
entire nuclear landscape. While the nuclear interaction  part of the nuclear EDF is constructed to be an isoscalar   \cite{(Per04),(Roh10)}, the Coulomb interaction breaks isospin manifestly. There are, therefore, two different sources of isospin symmetry breaking in the nuclear DFT: spontaneous isospin breaking associated with the self-consistent response to the neutron excess, and the explicit breaking due to the electric charge of the protons \cite{(Sat09)}. 

Effects related to isospin breaking and restoration are difficult to treat theoretically within the nuclear DFT. Below, we discuss two ways of dealing with this problem: isocranking and isospin projection.

\subsection{1D- and 3D-isocranking}
\label{sec:isocrank}

The isocranking model~\cite{(Sat03),(Sat06)}  attributes the kinetic coefficient $a_{\rm sym,kin}$ contribution to the
mean level spacing at the Fermi energy $\varepsilon (A)$ rather than to the total kinetic
energy itself.  The SHF calculations also revealed that the isovector mean potential
of the Skyrme EDF can be quite well characterized by an effective 
$V_{TT}$  interaction (\ref{vTT}) characterized by a strength parameter 
$\kappa (A)$.
The actual isovector part of the
Skyrme mean-field  potential  is composed of several terms  \cite{(Per04),(Roh10)}. As seen from Table~\ref{tab:sum-models}, in the uniform NM limit, two
terms contribute in SHF, $C_1^\rho \rho_1^2$  and $C_1^\tau \rho_1 \tau_1$,  and
the NSE strength reads:
\begin{equation}\label{inf}
   a_{\rm sym} = \frac{1}{8}\frac{m}{m^*} \varepsilon_{FG}
   + \left[   \left( \frac{3\pi^2}{2} \right)^{2/3} C_1^\tau \rho_0^{5/3}
   + C_1^\rho \rho_0 \right],
\end{equation}
where $\varepsilon_{\rm {FG}}$ is the average level splitting in FGM. Therefore, within this scenario, $a_{\rm sym}$ is
non-trivially modified by momentum-dependent effects
introducing, in the leading order, the dependence of $a_{\rm sym,kin}$ and
$a_{\rm sym,int}$ on the isoscalar and isovector effective mass, respectively.

Within the nuclear shell model, NSE appears through a 
contribution to the binding energy proportional to $T(T+1)$ \cite{(Tal62)}.
However, the local enhancement of binding around $N=Z$ (the Wigner energy) suggest an enhancement  of the linear term to $T(T+\lambda)$ with $\lambda\approx 1.26$ \cite{(Jan65),(Jan03),(Glo04)}. Since the Wigner energy 
is neither fully understood nor included properly within the SCMF models 
\cite{(Sat97)}, the microscopic origin of $\lambda$ is still a matter of debate.
Within the isocranking
model, the Fock exchange (isovector) potential gives rise to $\lambda \approx 0.5$, at variance with enhancement seen in experimental data. The Wigner energy
can be explained  by shell-model calculations \cite{(Sat97)} in terms of configuration mixing. The Wigner term
is usually associated with the isoscalar neutron-proton (np) pairing \cite{(Sat97a),(Roh10)}, but
its understanding is poor as realistic calculations
involving simultaneous np mixing in both the particle-hole (p-h) and particle-particle (p-p)
channels have not been carried out. It is only very recently that 3D isocranking calculations
including np mixing in the p-h channel have been reported \cite{(Sat13)}.
This is the first step towards developing the nuclear superfluid DFT including np mixing
in both p-h and p-p channels.  An improved  treatment of  isospin
within the 3D isocranking will open new opportunities for
quantitative studies of isobaric analogue states and, in turn, the NSE.

\subsection{Isospin projected DFT}
\label{sec:isoproj}

The isospin and isospin-plus-angular-momentum projected DFT models
have been developed recently to describe isospin mixing effects. These new tools open new avenues to probe NSE. To gain  insight  on this line of models, it is instructive to to consider the
spontaneous isospin symmetry breaking effect in the so-called anti-aligned
p-h configurations in $N = Z$ nuclei, which are mixtures of $T=0$ and $T=1$
states \cite{(Sat11a)}. Restoration of the isospin symmetry
results in the energy splitting, $\Delta E_T$, between the actual  $T=0$ and $T=1$ configurations.
Since these states are projected from a single mean-field determinant, the
splitting is believed to be insensitive to kinematics, and the method
can be used to probe dynamical effects giving rise to the interaction term
$a_{\rm sym,int}$. The results of SHF calculations \cite{(Sat11a)}
performed in finite nuclei confirm that $a_{\rm sym,int}$ is indeed correlated
with the isoscalar effective mass in agreement with the NM relation (\ref{inf}).

The isospin and isospin plus angular momentum projected DFT were designed and
applied to study the isospin impurities \cite{(Sat09)}
and isospin symmetry breaking corrections to the superallowed $0^+\rightarrow 0^+$
$\beta$-decay rates~\cite{(Sat11)}. Unfortunately, the calculations show that
these two observables are not directly correlated with the symmetry energy.
Ambiguities associated with these calculations stimulated further development
of the formalism in the direction of the Resonating-group method.
The scheme proceeds in three steps:
(i) 
First, a set of low-lying
(multi)p-(multi)h SHF states $\{ \Phi_i \}$
is calculated. These states
form a basis for a subsequent projection;
(ii)
Next,  the  projection techniques are applied
to calculate a family $\{ \Psi_I^{(\alpha )} \}$ of good angular momentum states
with properly treated $K$-mixing and isospin mixing;
(iii) 
Finally, a configuration mixing of $\{ \Psi_I^{(\alpha )} \}$ states is performed using techniques suitable for non-orthogonal ensambles.

\begin{figure}
{%
\centerline{\includegraphics[width=0.8\linewidth,clip=]{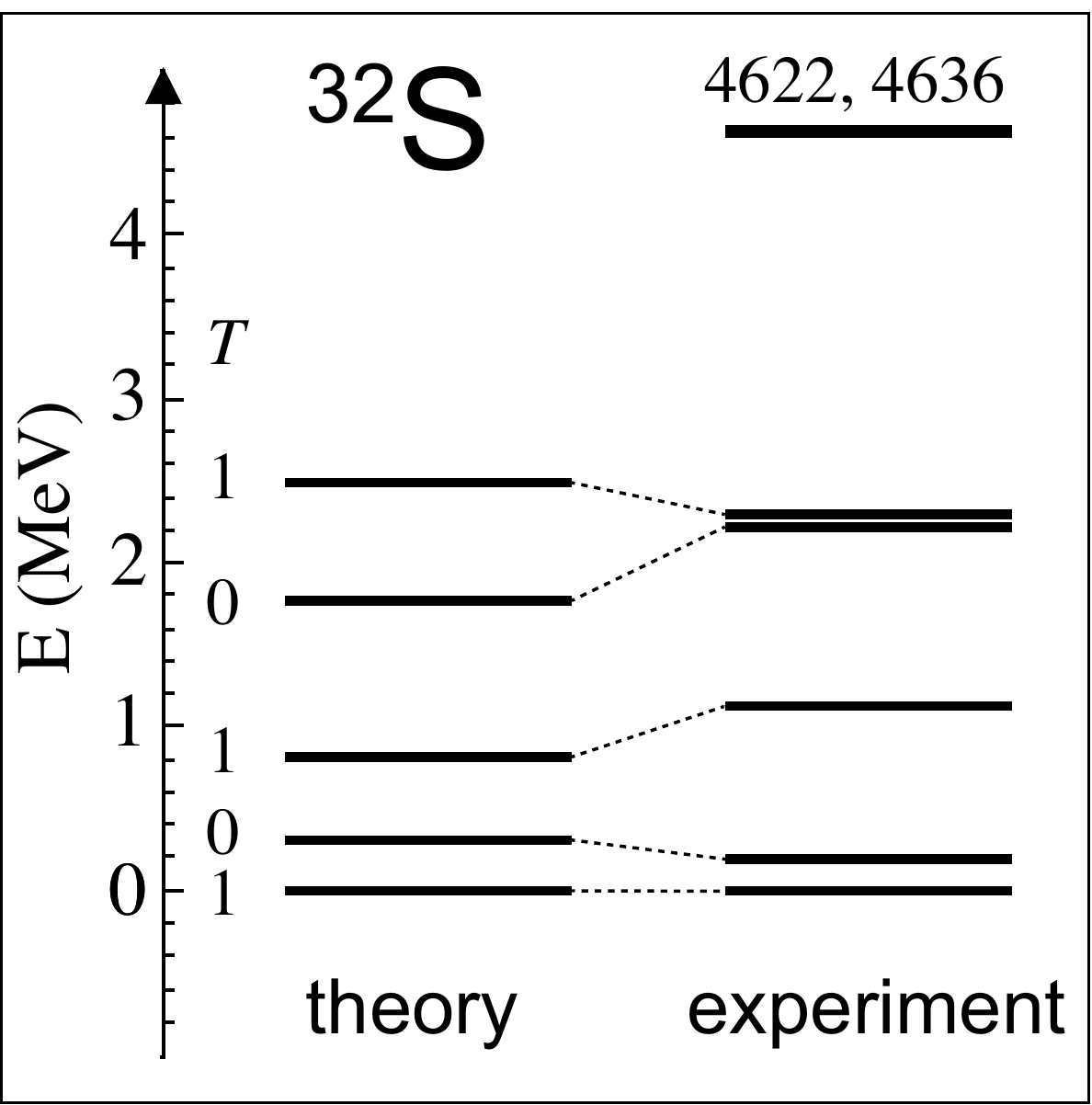}}
}
\caption{\label{fig:32s}
Energies of $I^\pi = 1^+$ states in $^{32}$S normalized to  the isobaric analogue state 
$I^\pi = 1^+, T=1$.
The results of projected
SHF-SkV calculations involving configuration mixing \cite{(Sat13a)} (left) are compared to experiment (right).
The calculations are based on 24 $I=1^+$ states projected from 6  HF determinants representing low-lying 1p-1h configurations.
}
\end{figure}

Although at present the calculations can be realized only for the SkV  EDF, the preliminary results~\cite{(Sat13a)} are encouraging, as shown in
Fig.~\ref{fig:32s}. Since the projected approach treats rigorously
the angular momentum conservation and the
long-range polarization due to the Coulomb force, it opens up a possibility
of detailed studies of the isovector terms of the nuclear EDF that are sources of the NSE.

\section{Conclusions}
\label{sec:concl}

This work surveys various aspects of NSE within the nuclear DFT  represented by
non-relativistic and relativistic self-consistent mean-field frameworks.
After defining the models and statistical tools, we reviewed  key observables pertaining to bulk nucleonic matter  and finite nuclei. Using the statistical covariance technique, constraints on the symmetry energy were studied, together with  correlations between observables and symmetry-energy parameters. 

Through the systematic correlation analysis, we scrutinized various observables
from finite nuclei that are accessible by current and future experiments. We confirm that by far the most sensitive isovector indicators are observables related to the neutron skin (neutron radius, diffraction radius, weak charge form factor) and the dipole polarizability \cite{Rei10,(Pie12)}. In this context, PREX-II measurement of the neutron skin in $^{208}$Pb \cite{Prex2} (a follow-up measurement to PREX \cite{(Abr12)}
designed to improve the experimental precision), CREX measurement of the neutron skin in $^{48}$Ca \cite{Crex}, and on-going measurements of $\alpha_{\rm D}$ in neutron-rich nuclei \cite{(Tam13)} are indispensable.

The masses of heavy neutron-rich nuclei also seem to correlate well with NSE parameters. Other observables, such as $Q_\alpha$-values, $\delta V_{pn}$, barrier heights, and low-energy dipole strength   \cite{Rei10,(Dao12),Rei13} are too strongly impacted by shell effects to be useful as {\it global} isovector indicators.

A  major challenge is to develop the universal nuclear EDF with improved isovector properties. 
Various improvements are anticipated in the near future. Those include constraining the EDF  at sub-saturation densities using ab initio models 
\cite{(Bog11),(Mar13)} and  
using the density matrix expansion  to develop an EDF based on microscopic nuclear interactions \cite{(Sto10)}. This work will be carried out under the Nuclear Low Energy Computational Initiative (NUCLEI) \cite{NUCLEI}.
Other exciting avenues are related to  multi-reference isospin projected DFT, which will enable us to make reliable predictions for isobaric analogues, isospin mixing, and mirror energy differences.

\begin{acknowledgement}
{\it This work was
supported by
the U.S. Department of Energy under
Contract No.\ DE-FG02-96ER40963
 (University of Tennessee), 
No. DE-SC0008499    (NUCLEI SciDAC Collaboration); by BMBF under Contract No. 06~ER~142D; and by NCN under Contract No. 2012/07/B/ST2/03907}
\end{acknowledgement}

%
%
%

\end{document}